\documentclass[12pt]{article}
\usepackage{pst-plot,epsf}
\newcommand{\beq}{\begin{equation}}
\newcommand{\eeq}{\end{equation}}
\newcommand{\bea}{\begin{eqnarray}}
\newcommand{\eea}{\end{eqnarray}}

\newcommand{\epm}{e^+e^-}

\newcommand{\ra}{\rightarrow}

\newcommand{\ttbar}{t\bar{t}}

\newcommand{\eebwbw}{e^+ e^- \ra b W^+ \bar{b} W^-}

\newcommand{\budbmn}{b u \bar{d} \bar{b} \mu^- \bar{\nu}_{\mu}}
\newcommand{\bnmbtn}{b \nu_{\mu} \mu^+ \bar{b} \tau^- \bar{\nu}_{\tau}}

\begin{document}
\thispagestyle{empty}
\begin{flushright}
April 2002\\
Revised version\\
July 2002\\
\vspace*{1.5cm}
\end{flushright}
\begin{center}
{\LARGE\bf {\tt eett6f v.~1.0}\\[4mm]
 A program for top quark pair production and decay 
           into 6 fermions at linear colliders\footnote{Work supported
           in part by the Polish State Committee for Scientific Research
           (KBN) under contract No. 2~P03B~004~18 and by the European 
           Community's Human Potential Programme under contract 
           HPRN-CT-2000-00149 Physics at Colliders.}}\\
\vspace*{2cm}
Karol Ko\l odziej\footnote{E-mail: kolodzie@us.edu.pl}\\[1cm]
{\small\it
Institute of Physics, University of Silesia\\ 
ul. Uniwersytecka 4, PL-40007 Katowice, Poland}\\
\vspace*{3.5cm}
{\bf Abstract}\\
\end{center}
The first version of a computer program {\tt eett6f} for calculating 
cross sections of $\epm \ra $ 6 fermions processes relevant for
a $\ttbar$-pair production and decay at centre of mass energies 
typical for linear colliders is presented.
{\tt eett6f v.~1.0} allows for calculating both the total and differential 
cross sections at tree level of the Standard Model (SM). 
The program can be used as the Monte Carlo generator of unweighted
events as well.
\vfill
\newpage
{\large \bf PROGRAM SUMMARY}\\[4mm]
{\it Title of program:} {\tt eett6f}

{\it Version:} 1.0 (April 2002)

{\it Catalogue identifier:} FJA0815b

{\it Program obtainable from:} CPC Program Library or on request by e-mail
from the author

{\it Licensing provisions:} none

{\it Computers:} all

{\it Operating systems:} Unix/Linux

{\it Programming language used:} {\tt FORTRAN 90}

{\it Memory required to execute with typical data:} 1 Mb

{\it No. of bits in a word:} 32

{\it No. of bytes in distributed program, including test data, etc.:} 400 Kb

{\it Distribution format:} ASCII

{\it Keywords:} $\epm$ annihilation, SM, lowest order six-fermion
reactions, $t\bar{t}$ production and decay

{\it Nature of physical problem}\\
Description of $\epm \ra $ 6 fermions processes relevant for
a $\ttbar$-pair production and decay at centre of mass energies 
typical for linear colliders; all six fermion reactions containing
a $b$ and $\bar{b}$ quarks and four other fermions of different flavours
with a complete set of the Feynman diagrams in the lowest order of SM.

{\it Method of solution}\\
Matrix elements are calculated numerically with the helicity amplitude
method. Constant widths of unstable particles are implemented by
modifying mass parameters in corresponding propagators. The phase space 
integration is performed numerically utilizing
a multi-channel Monte Carlo method.

{\it Restrictions on complexity of the problem}\\
Reactions containing fermions of the same flavour are not treated. No higher
order effects are taken into account, except for assuming the fine structure
constant and the strong coupling at appropriate scale and partial summation
of the one particle irreducible loop corrections by introducing fixed widths of
unstable particles.

{\it Typical running time}\\
The running time depends strongly on a desired precision of the result.
The results of the appended test run have been obtained on a 800~MHz
Pentium~III processor with the use of Absoft {\tt FORTRAN~90} compiler
in 490 seconds. In order to obtain a precision level below one per mille a few
million calls to the integrand are required. This results in a typical
running time of several hours. The running time becomes much shorter
for approximated cross sections.

\newpage

{\large \bf LONG WRITE-UP}\\
\section{Introduction}
Precise measurements of the top quark properties and interactions 
are planned at TESLA \cite{Tesla} and will most certainly belong to the 
research program of any future $\epm$ collider \cite{NLC}.
The measurements should be confronted with theoretical predictions
matching the same high precision level of a few per mille.
It is obvious that in order to reach that high precision it is mandatory
to include radiative corrections and, as the measurements 
of some top quark physical properties will be carried out at high energies, 
much above the $t\bar{t}$ threshold, it is crucial to know off-resonance 
background contributions to any specific 6 fermion decay channel and 
to estimate the effects related to the off-shellness of the $t\bar{t}$-pair. 

In this article, a technical documentation of a numerical program, 
{\tt eett6f}, is presented which allows
for computer simulation of the 6 fermion reactions which are relevant for 
the top quark pair production and its decay into a specific 
6 fermion final state
\beq
\label{eesixf}
  e^+(p_1) + e^-(p_2)\;\; \ra \;\; b(p_3) + f_1(p_4) + \bar{f'_1}(p_5) 
                               + \bar{b}(p_6) + f_2(p_7) + \bar{f'_2}(p_8),
\eeq
where $f_1=\nu_{\mu}, \nu_{\tau}, u, c$, $f_2=\mu^-, \tau^-, d, s$, and
$f'_1$, $f'_2$ are the corresponding weak isospin partners of $f_1$, $f_2$,
$f'_1=\mu^-, \tau^-, d, s$, $f'_2=\nu_{\mu}, \nu_{\tau}, u, c$, and the 
particle four momenta have been indicated in the parentheses.
In the present version, {\tt v.~1.0}, of the program, reactions
(\ref{eesixf}) are treated in the lowest order of SM.
Moreover, for the sake of simplicity, it is assumed that the 
actual values of $f_1$ and $f'_2$ in (\ref{eesixf}) are 
different from each other, and that neither $f'_1$ nor $f_2$ is an electron.
The physics contents of the program is described in \cite{KK}, where the
details on the physical model, a method of calculation and physical results 
obtained with {\tt eett6f v1.0} have been discussed. 

A similar analysis of the 6 fermion processes relevant for a $\ttbar$
production in $\epm$ annihilation have been performed
in \cite{ABP}, \cite{YKK}, where semileptonic channels
of reaction (\ref{eesixf}) have been studied, and
in \cite{Gangemi}, where 
purely hadronic channels of (\ref{eesixf}) have been analysed.
Moreover, irreducible QCD background to top searches in semileptonic channels
of (\ref{eesixf}) has been discussed in \cite{Moretti}.
After this work has already been completed an exstensive study of
six fermion reactions, including  (\ref{eesixf}) among others, in the 
massless fermion limit has appeared \cite{DR}. 
Using the same input parameters and separation cuts 
as in \cite{DR}, the present program gives 5.8160(32) fb and 17.223(15) fb
for the cross sections of $\epm \ra \bnmbtn$ and $\epm \ra \budbmn$,
respectively, at CMS 
energy of 500 GeV. These results are 2--3 standard deviations
off the corresponding results of \cite{DR}. This difference can most probably
be traced back to somewhat different implementation of the top quark
width in the top quark propagator. In the present work, the top quark width 
is introduced both in the denominator and in the numerator of the top 
propagator while, in \cite{DR}, it seems to be introduced in the denominator
only. The invariant mass and angular distributions of the $bu\bar{d}$
quark triple of $\epm \ra \budbmn$ at $s^{1/2}=500$ GeV shown in
\cite{DR} are nicely reproduced by {\tt eett6f} within accuracy  of the plots.

The main advantage of the present program is that it allows for taking 
into account both the electroweak and QCD lowest order contributions. 
As light fermion masses are not neglected, the cross sections
can be calculated without any kinematical cuts. Moreover, a number of 
options have been implemented in the program which make possible calculation
of the cross sections while switching on and off different subsets of the 
Feynman diagrams. It is possible to calculate cross sections in two different
narrow width approximations discussed below, too.

Besides of a possibility of taking into account solely the electroweak
contributions, or switching off the Higgs boson exchange diagrams,
the program allows also for a simplified treatment of reaction (\ref{eesixf}) 
by utilizing  a few different approximations:
the double resonance approximation for $W$ bosons
\bea
\label{doubleW}
  e^+(p_1) + e^-(p_2) & \ra &
      b(p_3) + {W^+}^*(p_{45}) + \bar{b}(p_6) + {W^-}^*(p_{78}) \\
&\ra &
\label{final1}
b(p_3) + f_1(p_4) + \bar{f'_1}(p_5) 
                               + \bar{b}(p_6) + f_2(p_7) + \bar{f'_2}(p_8),
\eea
where only those 61 Feynman diagrams are taken into account which contribute
to $\eebwbw$ and the W bosons are considered as being off-mass-shell and
the double resonance approximation for a $t$- and $\bar{t}$-quark
\bea
\label{eettsixf}         
  e^+(p_1) + e^-(p_2) & \ra &
      t^*(p_{345}) + \bar{t}^*(p_{678})  \\
\label{final2}
&\ra &
b(p_3) + f_1(p_4) + \bar{f'_1}(p_5) 
                               + \bar{b}(p_6) + f_2(p_7) + \bar{f'_2}(p_8),
\eea
with only two `signal' diagrams contributing. The intermediate state
momenta are $p_{45} = p_4 + p_5$, $p_{78} = p_7 + p_8$ in Eq.~(\ref{doubleW})
and $p_{345} = p_3 + p_4 + p_5$, $p_{678} = p_6 + p_7 + p_8$ in 
Eq.~(\ref{eettsixf}). The narrow width approximation
for the $W$ bosons (top quarks) is obtained by replacing the intermediate 
state $W$ bosons (top quarks) in Eq.~(\ref{doubleW}) (in Eq.~(\ref{eettsixf}))
 with on mass shell $W$'s (top quarks) and by multiplying the 
cross section of the on mass shell reaction (\ref{doubleW}) 
$\left((\ref{eettsixf})\right)$ with the branching ratios corresponding 
to final state of reaction (\ref{eesixf}). The necessary partial decay width
of the $W$ and $t$ are calculated according to the lowest order SM, and
the fermion masses are neglected in the three body top decay with.

The necessary matrix elements of reactions (\ref{eesixf}), (\ref{doubleW}) and
(\ref{eettsixf}) are calculated with the helicity amplitude method described 
in \cite{KZ} and \cite{JK} and phase space integrations are performed with 
the Monte Carlo (MC) method. More details on the multi-channel MC algorithm
(see e.g. \cite{MC}), on which the MC integration and event generation is 
based, are given in the next section.

\section{Monte Carlo integration and event generation}
In order to improve the
convergence of the MC integration the most relevant peaks of
the matrix element squared related to the Breit-Wigner shape of the 
$W, Z$, Higgs 
and top quark resonances as well as to the exchange of a massless photon or 
gluon have to be mapped away. 
As it is not possible to find out a single parametrization of the 
$n$-dimensional phase space which would allow to cover the whole resonance
structure of the integrand, it is necessary to utilize a multi-channel
MC approach. In the multi-channel MC approach, the random numbers 
$ 0 \le x_j \le 1$, $j=1,...,n$, and hence four momenta 
of the final state particles, are generated according to one of $N$ different 
probability densities $f_i(x)$, $i=1,...,N$, $x=(x_1,...,x_n)$. 
The distribution $f_i(x)$,
which accounts for several different peaks of the integrand,
is selected with probability $a_i$. All the distributions $f_i(x)$ are then 
combined in a single probability distribution $f(x)$ which should cover 
possibly all the peaks of the integrand, or at least 
the most relevant ones
\bea
\label{weights}
  f(x)=\sum_{i=1}^N a_i f_i(x), \qquad {\rm with} \qquad \sum_{i=1}^N a_i = 1.
\eea
The second condition of Eq.~(\ref{weights}) guaranties that, if every 
distribution
$f_i(x)$ is normalized to unity, then the combined distribution
$f(x)$ is normalized to unity, too. Obviously, all the distributions $f_i(x)$
and weights $a_i$, $i=1,...,N$, are to be non negative.

The basic phase space parametrizations which are used in the program are 
listed below. The 6 particle phase space of reaction (\ref{eesixf}) is
parametrized in 3 different ways:
\bea
\label{dps61}
 {\rm d}^{14} Lips  &=& (2\pi)^{-14}  
          {\rm d} PS_2\left(s,s_{345},s_{678}\right)
          {\rm d} PS_2\left(s_{345},m_3^2,s_{45}\right)
          {\rm d} PS_2\left(s_{678},m_6^2,s_{78}\right) \nonumber \\
&\times&  {\rm d} PS_2\left(s_{45},m_4^2,m_5^2\right)
          {\rm d} PS_2\left(s_{78},m_7^2,m_8^2\right)
          {\rm d} s_{345} {\rm d} s_{678} {\rm d} s_{45} {\rm d} s_{78},
\eea
\bea
\label{dps62}
 {\rm d}^{14} Lips  &=& (2\pi)^{-14}  
          {\rm d} PS_2\left(s,s_{34},s_{5678}\right)
          {\rm d} PS_2\left(s_{5678},s_{56},s_{78}\right)
          {\rm d} PS_2\left(s_{34},m_3^2,m_4^2\right) \nonumber \\
&\times&  {\rm d} PS_2\left(s_{56},m_5^2,m_6^2\right)
          {\rm d} PS_2\left(s_{78},m_7^2,m_8^2\right)
          {\rm d} s_{34} {\rm d} s_{5678} {\rm d} s_{56} {\rm d} s_{78}
\eea
and
\bea
\label{dps63}
 {\rm d}^{14} Lips  &=& (2\pi)^{-14}  
          {\rm d} PS_2\left(s,m_3^2,s_{45678}\right)
          {\rm d} PS_2\left(s_{45678},s_{45},s_{678}\right)
          {\rm d} PS_2\left(s_{678},m_6^2,s_{78}\right) \nonumber \\
&\times&  {\rm d} PS_2\left(s_{45},m_4^2,m_5^2\right)
          {\rm d} PS_2\left(s_{78},m_7^2,m_8^2\right)
          {\rm d} s_{45678} {\rm d} s_{45} {\rm d} s_{678} {\rm d} s_{78}.
\eea
The 4 particle phase space of reaction (\ref{doubleW}) in the narrow $W$ width
approximation is parametrized in 2 different ways:
\beq
\label{dps41}
 {\rm d}^{8} Lips  = (2\pi)^{-8}  
          {\rm d} PS_2\left(s,m_3^2,s_{456}\right)
          {\rm d} PS_2\left(s_{456},m_4^2,s_{56}\right) 
          {\rm d} PS_2\left(s_{56},m_5^2,m_6^2\right)
          {\rm d} s_{456} {\rm d} s_{56}
\eeq
and
\beq
\label{dps42}
 {\rm d}^{8} Lips = (2\pi)^{-8}  
          {\rm d} PS_2\left(s,s_{34},s_{56}\right)
          {\rm d} PS_2\left(s_{34},m_3^2,m_4^2,\right)
          {\rm d} PS_2\left(s_{56},m_5^2,m_6^2\right)
          {\rm d} s_{34} {\rm d} s_{56}.
\eeq
In Eqs.~(\ref{dps61}--\ref{dps42}), $s_{ijk...}=(p_i+p_j+p_k+...)^2,
i,j,k = 3,...,8$, and ${\rm d} PS_2\left(s,s',s''\right)$
is a two particle (subsystem) phase space element defined by
\beq
\label{dps2}
 {\rm d} PS_2\left(s,s',s''\right) = \delta^4\left( p - p' - p'' \right) 
   \frac{{\rm d}^3p'}{2E'} \frac{{\rm d}^3p''}{2E''} 
    = \frac{|\vec{p}\;'|}{4\sqrt{s}} {\rm d} \Omega',
\eeq
where $\vec{p}\;'$ is the momentum and $\Omega'$ is the solid angle
of one of the particles (subsystems) in the relative centre of mass system, 
$\vec{p}\;' + \vec{p}\;'' = 0$.
Making use of the rotational symmetry with respect to the $e^+e^-$ beam 
axis reduces the dimension of the phase space elements to 13 in 
Eqs.~(\ref{dps61})--(\ref{dps63}) and to 7 in Eqs.~(\ref{dps41}) and 
(\ref{dps42}).

Parametrizations (\ref{dps61}--\ref{dps42}) are used with different
permutations of external particle momenta such that invariants
$s_{ijk...}$ possibly correspond to the virtuality of 
propagators of the gauge bosons, Higgs boson and/or top quarks. The 
invariants are then transformed to the interval $[0,1]$ by performing
mappings smoothing out peaks related to the propagators. All other
integration variables of Eqs.~(\ref{dps61}--\ref{dps42}), not related
to peaks of the matrix element squared, are also transformed to the 
interval $[0,1]$ 
with a simple mapping defined in the following way. Let $y_j \in [c_j,b_j]$
be a variable not related to a peak in the matrix element squared and
$x_j$ be random variable uniformly distributed in the interval $[0,1]$,
then the relation
\beq
           y_j=\left(b_j - c_j\right) x_j + c_j
\eeq
defines the necessary mapping.
After having all the integration variables transformed 
to the interval $[0,1]$, any of phase space parametrizations 
(\ref{dps61}--\ref{dps63}) for reaction (\ref{eesixf}), or
(\ref{dps41}), (\ref{dps42}) for reaction (\ref{doubleW}) in the narrow
width approximation, for a given permutation of particle momenta, can 
be directly associated with the probability density $f_i(x)$, which is
referred to as a kinematical channel. Altogether 59 (21) different channels 
are used in order to integrate
the matrix element squared of reaction (\ref{eesixf}) (reaction 
(\ref{doubleW}) in the narrow width approximation).

The weights $a_i$ are calculated in the initial scanning run that starts
with all the weights equal to each other. The weights $a_i$, $i=1,...,N$,
for the actual run are calculated as ratios
\beq
\label{ai}
          a_i=\sigma_i/\sum_{j=1}^N \sigma_j,
\eeq
where $\sigma_j$ denotes the cross section obtained with the $i$-th 
kinematical channel in the initial scan.

Calculation of the cross section of reaction (\ref{eettsixf}) in the narrow
top width approximation does not require
the multi-channel MC approach and can be performed with the single
phase space parametrization (\ref{dps2}).

\section{Description of the program}
{\tt eett6f} is a package written in {\tt FORTRAN 90}. It consists of
50 files including a makefile. They are stored in one working directory.
The user should 
specify the physical input parameters in module {\tt inprms.f} and select 
a number of options in the main program {\tt csee6f.f}. 

\subsection{Program input}
The default values of the input parameters and options used in the program
are those specified below.
\subsubsection{Physical parameters}
The initial physical parameters to be specified in module {\tt inprms.f}
are the following.

Gauge boson masses and widths (GeV):\\[2mm]
{\tt  mw=80.419} GeV, the W mass,\\
{\tt  gamw=2.12} GeV, the W width,\\
{\tt  mz=91.1882} GeV, the Z mass,\\
{\tt  gamz=2.4952} GeV, the Z width.\\[2mm]
The electroweak (EW) mixing parameter {\tt sw2} is then calculated from\\[2mm]
\centerline{\tt sw2 = 1-mw2/mz2 }\\[2mm]
utilizing {\tt mw2 = mw**2}, {\tt mz2 = mz**2} in the fixed width scheme, and
\\[2mm]
{\tt mw2 = mw**2 - i*mw*gamw}, {\tt mz2 = mz**2 - i*mz*gamz} 
in the complex mass scheme.\\[2mm]
{\tt ralp0 = 137.03599976,} an inverse of the fine structure constant in 
the Thomson limit,\\[2mm]
{\tt gmu=1.16639}$\times {\tt 10}^{\tt -5}$ GeV$^{-2}$, the Fermi coupling 
constant,\\[2mm]
{\tt  alphas=0.1185}, the strong coupling constant at {\tt mz}.\\[2mm]
An inverse of the fine structure constant at {\tt mw}, {\tt ralpw}, is
then calculated from\\[2mm]
\centerline{\tt  ralpw=4.44288293815837/(2*sw2*gmu*mw**2).}\\[2mm]
{\tt  mh=115.0} GeV, the Higgs boson mass,\\[2mm]
{\tt  gamh=0} GeV, the Higgs boson width.\\[2mm]
If {\tt gamh = 0}, then the Higgs width is calculated according to
the lowest order of SM.\\[2mm]
Fermion masses and widths:\\[2mm]
{\tt me = 0.510998902} MeV, {\tt game = 0} MeV, for an electron,\\[2mm]
{\tt mmu = 105.658357} MeV, {\tt gammu = 0} MeV, for a muon,\\[2mm]
{\tt mtau=1.77703} GeV, {\tt gamtau = 0} GeV, for a lepton tau,\\[2mm]
{\tt mu = 5} MeV, {\tt gamu = 0} MeV, for an up quark,\\[2mm]
{\tt md=9} MeV, {\tt gamd = 0} MeV, for a down quark,\\[2mm]
{\tt mc=1.3} GeV, {\tt gamc = 0} GeV, for a charm quark,\\[2mm]
{\tt ms=150} MeV, {\tt gams = 0} MeV, for a strange quark,\\[2mm]
{\tt mt=174.3} GeV, {\tt gamt = 1.5} GeV, for a top quark,\\[2mm]
{\tt mb=4.4} GeV, {\tt gamb = 0} GeV, for a bottom quark.\\[4mm]
{\tt  ncol=3}, the number of colours,\\[2mm]
{\tt  convc=0.389379292}$\times 10^{12}$ fb ${\rm GeV}^2$, a 
conversion constant.
\subsubsection{Main options}
The following main options should be selected in the main program 
{\tt csee6f.f}.
                                                         \\[2mm]
The number of different centre of mass (CMS) energies {\tt ne}\\[2mm]
{\tt ne = 1}. {\em Recommended if the unweighted event are to be generated}.
                                     \\[2mm]
The actual values of the CMS energies in the array {\tt aecm} of size 
{\tt ne}:\\[2mm]
{\tt aecm=(/500.d0/)}.\\[2mm]
The final state of (\ref{eesixf}) by selecting a decay mode of 
the intermediate $W^+$, $W^-$ bosons, {\tt iwp, iwm = 1 or 2},
where 1 corresponds to a leptonic, and  2 to a hadronic decay mode,
and by specifying the family index {\tt ifp,ifm = 1,2,3} of the fermion
pair resulting from the $W^+$ or $W^-$ decay. For example, the reaction
$\epm \ra b \nu_{\mu}\mu^+ \bar{b} d \bar{u}$
corresponds to $W^+$ decaying leptonically and $W^-$ decaying hadronically
and the corresponding flags are:\\[2mm]
{\tt  iwp = 1}\\
{\tt  ifp = 2}\\
{\tt  iwm = 2}\\
{\tt  ifm = 1}.\\[2mm]
Whether or not to calculate the Born cross section utilizing the full 6 
particle kinematics, {\tt iborn = 1 (yes) / else (no)}, with {\tt ncall0}
calls to the integrand\\[2mm]
{\tt iborn = 1}\\
{\tt ncall0 = 20000}. {\em Recommended No of calls is a few millions.}\\[2mm]
Generate the unweighted events or not, {\tt imc = 1(yes)/else(no)}?\\[2mm]
{\tt  imc = 0.}\\[2mm]
No standard event record is used. If {\tt imc = 1}, then the final state 
particle momenta of the accepted unweighted events are printed in the output.
                                                                       \\[2mm]
Scan the Born cross section with {\tt nscan0} calls, {\tt iscan0 = 
1(yes)/else(no)}, in order to find the dominant kinematical channels, adjust
integration weights and find out the maximum value of the cross section\\[2mm]
{\tt  iscan0 = 1}.   {\em This option is strongly recommended.}\\
{\tt  nscan0 = 200}. {\em About 1 thousand is recommended.}\\[2mm]
Calculate an approximate cross section of (\ref{doubleW}), {\tt iwwa
= 1 (yes) / else (no)?}\\[2mm]
{\tt  iwwa = 1}\\
{\tt  ncallww = 20000}. {\em Recommended value is several hundred thousands.}
                                                                  \\[2mm]
Scan the Born cross section in the narrow W-width approximation with 
{\tt nscanww} calls, {\tt iscanww = 1(yes)/else(no)?}\\[2mm]
{\tt  iscanww = 1}. {\em Recommended.}\\
{\tt  nscanww = 200}. {\em Recommended value is about 1 thousand.}\\[2mm]
Calculate an approximate cross section of (\ref{eettsixf}), {\tt itopa = 
1(yes)/else(no)?}\\[2mm]
{\tt  itopa = 1}\\
{\tt  ncalla = 20000}.\\[2mm]
Include the Higgs boson exchange, {\tt ihiggs = 1(yes)/else(no)?}\\[2mm]
{\tt ihiggs = 1}\\[2mm]
Calculate the electroweak contributions only, {\tt iew = 
1(yes)/else(electroweak  + QCD \\ Feynman diagrams)?}\\[2mm]
{\tt iew = 0}.\\[2mm]
Calculate the $t\bar{t}$ resonant diagrams only, {\tt itt = 
1(yes)/else(all the diagrams)?}\\[2mm]
{\tt  itt = 0.}\\[2mm]
Calculate the double W resonant diagrams only, {\tt iww = 
1(yes)/else(all the diagrams)?}\\[2mm]
{\tt   iww = 0}.\\[2mm]
Choose the scheme: {\tt ischeme = 1(complex mass scheme)/else(fixed 
width scheme)}\\[2mm]
{\tt  ischeme = 1}.\\[2mm]
If {\tt ischeme=1}, then should {\tt alpha\_W} be complex {\tt (iaplw=1)} 
or real {\tt (ialpw=0)}?\\[2mm]
{\tt ialpw=1.}
\subsubsection{Auxiliary options}

Impose cuts, {\tt icuts = 1(yes)/else(no)?}\\[2mm]
{\tt  icuts = 0}\\[2mm]
If {\tt  icuts = 1}, then specify the kinematical cuts, {\em e.g.}:\\[2mm]
{\tt  ctlb = cos(10*pi/180)} -- cosine of the charged lepton--beam 
                                              angle,\\[2mm]
{\tt  ctqb = cos(5*pi/180)} -- cosine of the quark--beam angle,\\[2mm]
{\tt  ctll = 1} -- cosine of the charged lepton--charged lepton angle,
                                                                \\[2mm]
{\tt  ctlq = cos(5*pi/180)} -- cosine of the charged lepton--quark 
                                                    angle,\\[2mm]
{\tt  ecutl = 1} GeV -- minimum charged lepton energy,\\[2mm]
{\tt  ecutq = 3} GeV -- minimum quark energy,\\[2mm]
{\tt  mqq = 10} GeV -- minimum invariant mass of a quark pair,\\[2mm]
{\tt  mll = 0} GeV -- minimum invariant mass of a charged lepton pair, 
                                                                       \\[2mm]
Calculate distributions, {\tt idist = 1(yes)/0(no)?}\\[2mm]
{\tt  idis=0.}\\[2mm]
If {\tt idis = 1}, then specify parameters of the distributions:\\[2mm]
{\tt   xmin = (/ xl1, xl2, xl3, xl4/)} -- lower bounds,\\[2mm]
{\tt   xmax = (/ xu1, xu2, xu3, xu4/)} -- upper bounds,\\[2mm]
{\tt   nbs = (/ n1, n2, n3, n4/)} -- the corresponding numbers of bins 
in each distribution.\\[2mm]
Constants {\tt xli, xui} should be of type {\tt real(8)} and {\tt ni}
of type {\tt integer}. The number of desired distributions and the maximum
number of bins, {\tt nbmax = max\{n1, n2, n3, n4\}} should be 
specified in module {\tt distribs.f}.\\[2mm]
The maximum of the fully differential cross section {\tt crmax},
relevant only if {\tt iscan0 = 0} or {\tt iscanww = 0},\\[2mm]
{\tt  crmax=1000}.\\[2mm]

\subsection{Routines of {\tt eett6f}}

The main program {\tt csee6f.f}, each subroutine, function or module 
are located in a file named the same way as the routine itself, except
for 3 functions: {\tt srr, src and scc} located in a single file
{\tt dotprod.f}. The program consists of the following routines.

\begin{itemize}
\item {\tt SUBROUTINE} {\tt boost} -- returns a four vector boosted 
      to the Lorentz frame of velocity {\bf --v}.
\item {\tt SUBROUTINE} {\tt carlos} -- the MC integration routine.
\item {\tt SUBROUTINE} {\tt couplsmb} -- returns the SM couplings.
\item {\tt FUNCTION} {\tt cross} -- calculates the 
      cross section of (\ref{eesixf}), approximated cross sections of 
      (\ref{doubleW}--\ref{final2}) and 
      the cross section of (\ref{doubleW}) in the narrow $W$-width 
      approximation.
\item {\tt FUNCTION} {\tt crosstopa} -- calculates the cross 
      section of  (\ref{eettsixf}) in the narrow top width approximation.
\item The {\tt MAIN PROGRAM} {\tt csee6f}.
\item {\tt MODULE} {\tt distribs} -- contains parameters of distributions.
\item {\tt MODULE} {\tt drivec} -- contains driving flags and some kinematical 
      variables.
\item {\tt SUBROUTINE} {\tt eebwbw} -- returns the squared matrix element 
      averaged over initial spins and summed over final spins and colours of
      $\epm \ra bW^+\bar{b}W^-$.
\item {\tt SUBROUTINE} {\tt eee} -- returns a contraction of a triple gauge 
      boson coupling with three complex four vectors.
\item {\tt SUBROUTINE} {\tt eeee} -- returns a contraction of a quartic 
      gauge boson coupling with four complex four vectors.
\item {\tt SUBROUTINE} {\tt eeff1} -- returns the squared matrix element 
      averaged over initial spins and summed over final spins and colours of
      $\epm \ra f \bar{f}$.
\item {\tt SUBROUTINE} {\tt eeh} -- returns a contraction of the Higgs-gauge 
      boson coupling with two complex four vectors.
\item {\tt SUBROUTINE} {\tt eett6f} -- returns the squared matrix element 
      averaged over initial spins and summed over final spins and colours of
      reaction (\ref{eesixf}) and approximations 
      (\ref{doubleW}--\ref{final2}).
\item {\tt SUBROUTINES} {\tt eev, eve} and {\tt vee} -- return contractions
      of a triple gauge boson coupling with two polarization vectors
      leaving, respectively, the third, second and first Lorentz index 
      uncontracted.
\item {\tt SUBROUTINE} {\tt fhfb} -- returns a set complex scalars:
      $scal(\lambda_1,\lambda_2)=g_{S12}\bar{u_1}u_2 \Delta_F$,
      where $u_1$, $u_2$ are fermion spinors, $\Delta_F$ is a scalar
      boson propagator and $g_{S12}$ is a coupling of the scalar boson
      to fermion 1 and 2.
\item {\tt SUBROUTINE} {\tt fvfa} -- returns a set of four vectors
      $\varepsilon^{\mu}(\lambda_1,\lambda_2)=\bar{u}_1(\lambda_1)
      \gamma_{\nu}(g_V^{(-)}P_-  + g_V^{(+)}P_+)$ $\times u_2(\lambda_2)
      D_V^{\nu\mu}$, where $u_1$, $u_2$ are fermion spinors, $P_{\pm}=
      (1 \pm \gamma_5)/2$, are chirality projectors, $g_V^{(\pm)}$ are
      the fermion-gauge boson couplings of definite chirality and
      $D_V^{\nu\mu}$ is the propagator of a gauge boson $V$ in arbitrary
      linear gauge.
\item {\tt SUBROUTINE} {\tt fef} -- returns a set of matrix elements
      $mat(\lambda_1,\lambda,\lambda_2)=\bar{u_1}(\lambda_1)
      /\!\!\!\varepsilon(\lambda)(g_V^{(-)}P_-  + g_V^{(+)}P_+) 
      u_2(\lambda_2)$, where notation is the same as in {\tt fvfa}.
\item {\tt SUBROUTINE} {\tt impcuts} -- imposes cuts on the kinematical 
      variables of process (\ref{eesixf}).
\item {\tt MODULE} {\tt inprms} -- contains initial input parameters.
\item {\tt MODULE} {\tt kincuts} -- contains kinematical cuts.
\item {\tt SUBROUTINE} {\tt kinee1212} -- returns the four momenta, phase 
      space normalization and flux factor for a 2 $\ra$ 6 process in 
      the CMS. The phase space parametrization used is that of 
      Eq.~(\ref{dps63}).
\item {\tt SUBROUTINE} {\tt kinee6f} -- returns the four momenta, phase 
      space normalization
      and flux factor for a 2 $\ra$ 6 process in the CMS. The phase space
      parametrization used is that of Eq.~(\ref{dps62}).
\item {\tt SUBROUTINE} {\tt kineeff1} -- returns the four momenta, phase 
      space normalization and flux factor for a $2 \ra 2$ process in the CMS.
      The phase space is parametrized according to Eq.~(\ref{dps2}).
\item {\tt SUBROUTINE} {\tt kineeffff} -- returns the four momenta, phase 
      space normalization and flux factor for a 2 $\ra$ 4 process in 
      the CMS. The phase space is parametrized according to Eq.~(\ref{dps42}).
\item {\tt SUBROUTINE} {\tt kineetbff} -- returns the four momenta, phase 
      space normalization and flux factor for a 2 $\ra$ 4 process in the CMS. 
      The phase space is parametrized according to Eq.~(\ref{dps41}).
\item {\tt SUBROUTINE} {\tt kineett6f} -- returns the four momenta, phase 
      space normalization and flux factor for a 2 $\ra$ 6 process in the CMS. 
      The phase space parametrization used is that of Eq.~(\ref{dps61}).
\item {\tt SUBROUTINE} {\tt kinff} -- returns the final state four momenta 
      and phase space normalization for a $2 \ra 2$ process in the CMS.
\item {\tt FUNCTION} {\tt lamsq} -- the kinematic lambda function, 
      $\lambda(\sqrt{x},\sqrt{y},\sqrt{z})$.
\item {\tt MODULE} {\tt mathprms} -- contains necessary arithmetical constants.
\item {\tt SUBROUTINE} {\tt parfix} -- returns parameters of a specific 
      process (\ref{eesixf}) and the branching ratios for
      the narrow width approximations; initializes weights for the
      Monte Carlo integration.
\item {\tt MODULE} {\tt parproc} -- contains parameters for a specific process.
\item {\tt SUBROUTINE} {\tt peu} -- returns a set of generalized spinors:
      $u(\lambda,\lambda_1)=S_F /\!\!\!\varepsilon(\lambda)
      (g_V^{(-)}P_-  + g_V^{(+)}P_+)u_1(\lambda_1)$,
      where $S_F$ is a Feynman propagator of an internal fermion and the
      remaining notation is the same as in {\tt fvfa}.
\item {\tt SUBROUTINE} {\tt psnee1212} -- returns a phase space normalization
      as the one of {\tt kinee1212} for a given set of external particle 
      momenta.
\item {\tt SUBROUTINE} {\tt psnee6f} -- returns a phase space normalization
      as the one of {\tt kinee6f} for a given set of external particle 
      momenta.
\item {\tt SUBROUTINE} {\tt psneeffff} -- returns a phase space normalization
      as the one of {\tt kineeffff} for a given set of external particle 
      momenta.
\item {\tt SUBROUTINE} {\tt psneetbff} -- returns a phase space normalization
      as the one of {\tt kineetbff} for a given set of external particle 
      momenta.
\item {\tt SUBROUTINE} {\tt psneett6f} -- returns a phase space normalization
      as the one {\tt kineett6f} for a given set of external particle 
      momenta.
\item {\tt SUBROUTINE} {\tt recpol} -- returns real polarization vectors of
      a vector boson in the rectangular basis.
\item {\tt FUNCTION} {\tt scc} (contained in {\tt dotprod.f}) -- returns 
      the Minkowski dot product of two complex four vectors.
\item {\tt SUBROUTINE} {\tt spheric} -- returns spherical components {\tt ps}
       of a four vector $p^{\mu}$, {\tt ps}$=\\
       (|{\bf p}|,\cos{\theta},
       \sin{\theta},\cos{\phi},\sin{\phi})$, with $\theta$ and $\phi$ being 
       a polar and azimuthal angles of momentum {\bf p}.
\item {\tt SUBROUTINE} {\tt spinc} -- returns the contractions:
      $p^0 I - {\bf p \cdot \sigma}$ and $p^0 I + {\bf p \cdot \sigma}$,
      where $p^{\mu}=(p^0,{\bf p})$ is a complex four vector, $I$ is the 
      $2 \times 2$ unit matrix and $\sigma$ are the Pauli matrices.
\item {\tt SUBROUTINE} {\tt spinornew} -- returns helicity spinors in the Weyl 
      representation.
\item {\tt SUBROUTINE} {\tt spinr} -- returns the same contractions as
      in {\tt spinc} for a real four vector $p^{\mu}$.
\item {\tt FUNCTION} {\tt src} (contained in {\tt dotprod.f}) -- returns 
      the Minkowski dot product of a real and a complex four vector.
\item {\tt FUNCTION} {\tt srr} (contained in {\tt dotprod.f}) -- returns 
      the Minkowski dot product of two real four vectors.
\item {\tt SUBROUTINE} {\tt uep} -- returns a set of generalized spinors:
      $u(\lambda_1,\lambda)=u_1(\lambda_1) /\!\!\!\varepsilon(\lambda)
      (g_V^{(-)}P_-  + g_V^{(+)}P_+) S_F$,
      where notation is the same as in {\tt peu}.
\item {\tt SUBROUTINE} {\tt wwidth} -- returns the partial width of the 
      $W$-boson averaged over initial spins and summed over final spins, 
      and colours for hadronic decay modes.
\end{itemize}

\subsection{Run output}
A sample of the listing of the test run output is given in Appendix A.
It contains a specification of the process, information on the scheme choice, 
and values of the relevant physical parameters in the very beginning. 
Then, for each value of the CMS energy and for each cross section calculated,
results of the initial scan and resulting weights for the actual run
are printed, if {\tt iscan = 1}. Sometimes the weights calculated according to 
Eq.~(\ref{ai}) do not add up exactly to 1. Then, the last weight is 
changed slightly in order the second condition of Eq.~(\ref{weights}) to
be fulfilled. This change is supposed to be completely irrelevant numerically,
but the message about it is printed. The final result for the total cross 
section is called {\tt Integral}. Its value is printed together
with the standard deviation and the actual number of calls used in the 
calculation. Finally,
information on events acceptance efficiency is given, that means a fraction
of accepted weight 1 events. If the event generation
option is switched on, i.e. {\tt imc = 1}, then the corresponding
number of unwighted events will be printed out as collections of final
state particle four momenta.

Whenever a maximum value of the cross section initially assigned in 
{\tt csee6f.f} or found in the result of the initial scan is overflown by 
more than a factor 1.5, a corresponding message informing about it is printed. 
The integration is still valid, however, if the program is run as
an event generator, it should be rerun. How to proceed in this situation is 
described in the next section.
In the very end, all the calculated total cross sections together with 
the corresponding standard deviations are collected in the tabular form.
\section{Use of the program}
Up to now the program have been run only on Unix/Linux platforms.
In order to run the program, the user should select 
a specific name of a {\tt FORTRAN 90/95} compiler, desired options and the 
name of the output file in the {\tt makefile}. 
The output file is called {\tt test} at present.
The program can be then compiled, linked and run by executing a single
command\\[2mm]
{\tt make test}.\\[2mm]
The results of the test run should reproduce those contained
in file {\tt testrun}.

The program can be run as the MC event generator of unweighted
events by selecting \\[2mm]
{\tt imc = 1}\\[2mm]
in {\tt csee6f.f}. It is then recommended to run the program for a single CMS
energy and to perform a scan with a relatively large number of calls,
{\tt nscan0} or {\tt nscanww}. Attention should be paid to possible messages 
informing about updates of the maximum weight. In this case the program 
should be rerun, however, this time without the initial scan. The initial
integration weights {\tt aw0} or {\tt awa} in {\tt parfix.f}
and the maximum weight {\tt crmax} 
in {\tt csee6f.f} should be updated according to the results of the
prior scan. At present the efficiency of events acceptance is relatively
low. However, in view of a relatively fast performance of the program,
this should not be a serious limitation for the user.
\section{Outlook}
The following improvements of the program are envisaged in the near future.
First of all, a new version is being prepared which will allow 
for calculations of reactions involving the same external fermion flavours. 
Another step will be an
inclusion of the most relevant factorizable higher order effects: 
electroweak and QCD corrections to the top pair production, 
$e^+e^- \ra t \bar{t}$, to the $t$ and $\bar{t}$ quark decays, as well
as to the $W^+$ and $W^-$ decays.


%

\vspace*{1.5cm}


\centerline{\Large\bf {Appendix}}

\appendix
\section{Listing of test run output}
{\small \tt
\begin{center}
------------------------------------------------------------------------------
\\[2mm]
 Approaching the process:
\\[2mm]
                         e+ e- -> b vm mu+ b~ d u~  
\\[2mm]
 with the program 'eett6f v. 1.0' by
\\[2mm]
                            Karol Kolodziej
\\
                         University of Silesia
\\
                            Katowice, Poland
\\
                        e-mail: kolodzie@us.edu.pl
\\[2mm]
------------------------------------------------------------------------------
\\[2mm]
\end{center}
 Complex mass scheme
\\
 with sw2 = ( 0.222271237327414,-0.000778147496385)
\\
 and complex 1/alpha\_W = (132.386398893114400,  3.953796222427256)
\\[2mm]
 Gauge boson masses and widths: W, Gm\_W, Z, Gm\_Z
\\
     80.41900     2.12000    91.18820     2.49520
\\[2mm]
 Higgs boson mass and width: H, Gm\_H
\\
     0.11500E+03    0.49657E-02
\\[2mm]
 1/alpha0 = 137.0359998   alphas = 0.1185000   G\_mu =  0.116639E-04
\\[2mm]
 t-quark mass and width:
\\
    174.30     1.50000
\\[2mm]
 External fermion masses: e- b vm mu+ d u~  :
\\
 0.51099890E-03 0.44000000E+01 0.00000000E+00 0.10565836E+00
\\
 0.90000000E-02 0.50000000E-02
\\[2mm]
------------------------------------------------------------------------------
\\
\centerline{                       E\_cm =     500.000 GeV}
\\
------------------------------------------------------------------------------
\\
 All cross sections in femtobarns (fb)
\\
------------------------------------------------------------------------------
\\
 The Born cross section with full 6 particle kinematics:
\\
------------------------------------------------------------------------------
\\
 Results of the initial scan:
\\
 Kinematics        Weight          Integral        St.dev.       No of calls
\\
     1        0.17241379E-01   0.77943026E+01   0.53027331E+00       200
\\
     2        0.17241379E-01   0.70433588E+00   0.14557680E+00       200
\\
     3        0.17241379E-01   0.46787757E+00   0.14942639E+00       200
\\
     4        0.17241379E-01   0.40042444E-01   0.17572985E-01       200
\\
     5        0.17241379E-01   0.61761884E-01   0.17401289E-01       200
\\
     6        0.17241379E-01   0.70179588E-02   0.50890185E-02       200
\\
     7        0.17241379E-01   0.18918195E-04   0.14788334E-04       200
\\
     8        0.17241379E-01   0.59135660E-02   0.26123591E-02       200
\\
     9        0.17241379E-01   0.65946999E-05   0.37561359E-05       200
\\
    10        0.17241379E-01   0.71059323E-02   0.32477486E-02       200
\\
    11        0.17241379E-01   0.12331601E+00   0.52550633E-01       200
\\
    12        0.17241379E-01   0.50746197E-01   0.85455398E-02       200
\\
    13        0.17241379E-01   0.52341207E-01   0.13094625E-01       200
\\
    14        0.17241379E-01   0.16697241E-02   0.91218523E-03       200
\\
    15        0.17241379E-01   0.21727387E-02   0.96316906E-03       200
\\
    16        0.17241379E-01   0.10069801E+00   0.44469111E-01       200
\\
    17        0.17241379E-01   0.55418681E-01   0.10461875E-01       200
\\
    18        0.17241379E-01   0.47690194E-03   0.25400283E-03       200
\\
    19        0.17241379E-01   0.12270501E-05   0.68571740E-06       200
\\
    20        0.17241379E-01   0.57639038E-02   0.20367189E-02       200
\\
    21        0.17241379E-01   0.10948697E-03   0.78871872E-04       200
\\
    22        0.17241379E-01   0.38725698E-03   0.32739976E-03       200
\\
    23        0.17241379E-01   0.31913662E-03   0.31771241E-03       200
\\
    24        0.17241379E-01   0.18057468E-04   0.15825867E-04       200
\\
    25        0.17241379E-01   0.75726379E-02   0.10353862E-02       200
\\
    26        0.17241379E-01   0.40524491E-01   0.40417673E-01       200
\\
    27        0.17241379E-01   0.48346637E-03   0.25434560E-03       200
\\
    28        0.17241379E-01   0.17689713E-05   0.10354945E-05       200
\\
    29        0.17241379E-01   0.24943550E-03   0.85703152E-04       200
\\
    30        0.17241379E-01   0.68130329E+01   0.55624830E+00       200
\\
    31        0.17241379E-01   0.80133941E+00   0.14035567E+00       200
\\
    32        0.17241379E-01   0.37030192E+00   0.13116529E+00       200
\\
    33        0.17241379E-01   0.27961633E-01   0.79658869E-02       200
\\
    34        0.17241379E-01   0.12381885E-02   0.65639060E-03       200
\\
    35        0.17241379E-01   0.38608372E-05   0.23584221E-05       200
\\
    36        0.17241379E-01   0.22584220E-03   0.16055018E-03       200
\\
    37        0.17241379E-01   0.10262033E-05   0.42675278E-06       200
\\
    38        0.17241379E-01   0.59892263E-04   0.17034981E-04       200
\\
    39        0.17241379E-01   0.11805922E-04   0.96111792E-05       200
\\
    40        0.17241379E-01   0.37482281E-02   0.13933930E-02       200
\\
    41        0.17241379E-01   0.69961565E-02   0.29473064E-02       200
\\
    42        0.17241379E-01   0.26317460E-03   0.96401737E-04       200
\\
    43        0.17241379E-01   0.14981430E-04   0.14211747E-04       200
\\
    44        0.17241379E-01   0.36580810E-02   0.31164792E-02       200
\\
    45        0.17241379E-01   0.65694693E-05   0.59639961E-05       200
\\
    46        0.17241379E-01   0.59283100E-01   0.19915949E-01       200
\\
    47        0.17241379E-01   0.34966532E-02   0.32846811E-02       200
\\
    48        0.17241379E-01   0.39467871E-01   0.24414577E-01       200
\\
    49        0.17241379E-01   0.33496371E-02   0.76289358E-03       200
\\
    50        0.17241379E-01   0.65454536E+00   0.11507542E+00       200
\\
    51        0.17241379E-01   0.11775709E-01   0.58766647E-02       200
\\
    52        0.17241379E-01   0.12389848E-01   0.28548786E-02       200
\\
    53        0.17241379E-01   0.19047822E-01   0.13103839E-01       200
\\
    54        0.17241379E-01   0.29502206E-01   0.16858165E-01       200
\\
    55        0.17241379E-01   0.62510122E+00   0.13584265E+00       200
\\
    56        0.17241379E-01   0.82405870E-02   0.22527034E-02       200
\\
    57        0.17241379E-01   0.55814360E-02   0.23231953E-02       200
\\
    58        0.17241379E-01   0.78977140E-02   0.36720050E-02       200
\\
------------------------------------------------------------------------------
\\
 Born scanned with   11565 calls =  0.19039197E+02
\\
 The maximum value of the cross section=  0.27872147E+04
\\
 shifted to  0.41808220E+04
\\
------------------------------------------------------------------------------
\\
 Weight   58 =  0.4148134056762879E-03 rescaled to 0.4148134056756181E-03
\\
 Calculating the Born cross section with the weights:
\\
  1  0.40938191E+00  0.40938191E+00
\\
  2  0.36993992E-01  0.44637590E+00
\\
  3  0.24574439E-01  0.47095034E+00
\\
  4  0.21031583E-02  0.47305350E+00
\\
  5  0.32439333E-02  0.47629743E+00
\\
  6  0.36860583E-03  0.47666604E+00
\\
  7  0.99364459E-06  0.47666703E+00
\\
  8  0.31059956E-03  0.47697763E+00
\\
  9  0.34637490E-06  0.47697798E+00
\\
 10  0.37322648E-03  0.47735120E+00
\\
 11  0.64769547E-02  0.48382816E+00
\\
 12  0.26653539E-02  0.48649351E+00
\\
 13  0.27491290E-02  0.48924264E+00
\\
 14  0.87699297E-04  0.48933034E+00
\\
 15  0.11411925E-03  0.48944446E+00
\\
 16  0.52889841E-02  0.49473344E+00
\\
 17  0.29107678E-02  0.49764421E+00
\\
 18  0.25048428E-04  0.49766926E+00
\\
 19  0.64448631E-07  0.49766932E+00
\\
 20  0.30273882E-03  0.49797206E+00
\\
 21  0.57506089E-05  0.49797781E+00
\\
 22  0.20339986E-04  0.49799815E+00
\\
 23  0.16762084E-04  0.49801491E+00
\\
 24  0.94843646E-06  0.49801586E+00
\\
 25  0.39773936E-03  0.49841360E+00
\\
 26  0.21284769E-02  0.50054208E+00
\\
 27  0.25393213E-04  0.50056747E+00
\\
 28  0.92912076E-07  0.50056757E+00
\\
 29  0.13101157E-04  0.50058067E+00
\\
 30  0.35784246E+00  0.85842313E+00
\\
 31  0.42088930E-01  0.90051206E+00
\\
 32  0.19449451E-01  0.91996151E+00
\\
 33  0.14686352E-02  0.92143014E+00
\\
 34  0.65033652E-04  0.92149518E+00
\\
 35  0.20278362E-06  0.92149538E+00
\\
 36  0.11861960E-04  0.92150724E+00
\\
 37  0.53899508E-07  0.92150729E+00
\\
 38  0.31457348E-05  0.92151044E+00
\\
 39  0.62008511E-06  0.92151106E+00
\\
 40  0.19686903E-03  0.92170793E+00
\\
 41  0.36746070E-03  0.92207539E+00
\\
 42  0.13822779E-04  0.92208921E+00
\\
 43  0.78687301E-06  0.92209000E+00
\\
 44  0.19213421E-03  0.92228213E+00
\\
 45  0.34504971E-06  0.92228248E+00
\\
 46  0.31137396E-02  0.92539622E+00
\\
 47  0.18365550E-03  0.92557987E+00
\\
 48  0.20729799E-02  0.92765285E+00
\\
 49  0.17593374E-03  0.92782879E+00
\\
 50  0.34378833E-01  0.96220762E+00
\\
 51  0.61849819E-03  0.96282612E+00
\\
 52  0.65075477E-03  0.96347687E+00
\\
 53  0.10004530E-02  0.96447733E+00
\\
 54  0.15495510E-02  0.96602688E+00
\\
 55  0.32832332E-01  0.99885921E+00
\\
 56  0.43282221E-03  0.99929203E+00
\\
 57  0.29315502E-03  0.99958519E+00
\\
 58  0.41481341E-03  0.10000000E+01
\\
 59  0.00000000E+00  0.00000000E+00
\\
 60  0.00000000E+00  0.00000000E+00
\\
------------------------------------------------------------------------------
\\
 Final result:
\\
 Integral = 0.201003E+02  Standard dev. = 0.27E+00 with     19980 calls
\\
------------------------------------------------------------------------------
\\
 Efficiency of the events acceptance = 0.410E-02
\\
------------------------------------------------------------------------------
\\
 The narrow W-width approximation, e+ e- -> b W+ b~ W- -> 6f:
\\
------------------------------------------------------------------------------
\\
 Results of the initial scan:
\\
 Kinematics        Weight          Integral        St.dev.       No of calls
\\
     1        0.47619048E-01   0.23618560E+00   0.10375292E+00       200
\\
     2        0.47619048E-01   0.25574899E+00   0.11393103E+00       200
\\
     3        0.47619048E-01   0.90318927E-02   0.27624478E-02       200
\\
     4        0.47619048E-01   0.14178256E+00   0.63376575E-01       200
\\
     5        0.47619048E-01   0.29188135E-02   0.15778093E-02       200
\\
     6        0.47619048E-01   0.75109770E-02   0.28446236E-02       200
\\
     7        0.47619048E-01   0.87010486E-01   0.36285359E-01       200
\\
     8        0.47619048E-01   0.20008583E-01   0.10505322E-01       200
\\
     9        0.47619048E-01   0.25432033E+00   0.12331680E+00       200
\\
    10        0.47619048E-01   0.14796519E+00   0.51432260E-01       200
\\
    11        0.47619048E-01   0.14443649E+00   0.47481858E-01       200
\\
    12        0.47619048E-01   0.13431004E+00   0.57817377E-01       200
\\
    13        0.47619048E-01   0.62963498E-01   0.18297859E-01       200
\\
    14        0.47619048E-01   0.14057286E+00   0.83384476E-01       200
\\
    15        0.47619048E-01   0.16790199E+02   0.67749184E+00       200
\\
    16        0.47619048E-01   0.61926860E-01   0.12581699E-01       200
\\
    17        0.47619048E-01   0.12722393E+00   0.37044922E-01       200
\\
    18        0.47619048E-01   0.15741819E+00   0.64646409E-01       200
\\
    19        0.47619048E-01   0.16399987E+00   0.54156119E-01       200
\\
    20        0.47619048E-01   0.21743058E+00   0.13237257E+00       200
\\
    21        0.47619048E-01   0.13765610E+00   0.84254619E-01       200
\\
------------------------------------------------------------------------------
\\
 Born scanned with    4200 calls =  0.19300621E+02
\\
 The maximum value of the cross section=  0.98788719E+03
\\
 shifted to  0.14818308E+04
\\
------------------------------------------------------------------------------
\\
 Weight   21 =  0.7132210911535855E-02 rescaled to 0.7132210911536152E-02
\\
 Calculating the Born cross section with the weights:
\\
  1  0.12237203E-01  0.12237203E-01
\\
  2  0.13250817E-01  0.25488019E-01
\\
  3  0.46795867E-03  0.25955978E-01
\\
  4  0.73460104E-02  0.33301989E-01
\\
  5  0.15122900E-03  0.33453218E-01
\\
  6  0.38915728E-03  0.33842375E-01
\\
  7  0.45081704E-02  0.38350545E-01
\\
  8  0.10366808E-02  0.39387226E-01
\\
  9  0.13176795E-01  0.52564021E-01
\\
 10  0.76663435E-02  0.60230365E-01
\\
 11  0.74835153E-02  0.67713880E-01
\\
 12  0.69588457E-02  0.74672726E-01
\\
 13  0.32622525E-02  0.77934978E-01
\\
 14  0.72833337E-02  0.85218312E-01
\\
 15  0.86993051E+00  0.95514882E+00
\\
 16  0.32085424E-02  0.95835737E+00
\\
 17  0.65917015E-02  0.96494907E+00
\\
 18  0.81561205E-02  0.97310519E+00
\\
 19  0.84971292E-02  0.98160232E+00
\\
 20  0.11265471E-01  0.99286779E+00
\\
 21  0.71322109E-02  0.10000000E+01
\\
 The maximum weight updated from 0.14818308E+04 to  0.16809386E+04
\\
 The maximum weight updated from 0.16809386E+04 to  0.17190927E+04
\\
------------------------------------------------------------------------------
\\
 Final result:
\\
 Integral = 0.206457E+02  Standard dev. = 0.18E+00 with     20000 calls
\\
------------------------------------------------------------------------------
\\
 Efficiency of the events acceptance = 0.129E-01
\\
------------------------------------------------------------------------------
\\
 Approximated cross section e+ e- -> t t~ -> 6f:
\\
------------------------------------------------------------------------------
\\
 Final result:
\\
 Integral = 0.193058E+02  Standard dev. = 0.66E-01 with     20000 calls
\\
------------------------------------------------------------------------------
\\
 Born cross sections (fb):
\\
------------------------------------------------------------------------------
\\
    ecm         cs        sd        cstta     sda      csbbww     sdbbww
\\
 0.5000E+03 0.20100E+02 0.27E+00 0.19306E+02 0.66E-01 0.20646E+02 0.18E+00
\\
}


\begin{thebibliography}{99}
\bibitem{Tesla} TESLA Technical Design Report, Part III: Physics at an 
                $\epm$ Linear Collider, edited by
                R.-D. Heuer, D. Miller, F. Richard, P.M. Zerwas, 
                DESY 2001-011, ECFA 2001-209, TESLA Report 2001-23, 
                TESLA-FEL 2001-05, March 2001, hep-ph/0106315;
                $\epm$ Collisions at TeV Energies: The Physics Potential, 
                Proceedings of the Workshop, Annecy, Gran Sasso, Hamburg, 
                February 1995 -- September 1995, edited by P.M. Zerwas, 
                DESY 96--123D, (1996) 1.
\bibitem{NLC} T.~Abe {\it et al.}, American Linear Collider Working Group 
              Collaboration,
              SLAC-R-570, {\it Resource book for Snowmass 2001}.\\
              K.~Abe {\it et al.}, hep-ph/0109166.
\bibitem{KK} K. Ko\l odziej, Eur. Phys. J. C23 (2002) 471, [hep-ph/0110063].
\bibitem{ABP} E. Accomando, A. Ballestrero, M. Pizzio, Nucl.Phys. B512 (1998)
              19.
\bibitem{YKK} F.~Yuasa, Y.~Kurihara, S. Kawabata, Phys. Lett. B414 (1997)
              178.
\bibitem{Gangemi} F. Gangemi, G. Montagna, M. Moretti, O. Nicrosini,
                  F. Piccinini,  Nucl. Phys. B559 (1999) 3.
\bibitem{Moretti} S. Moretti, Eur. Phys. J. C9 (1999) 229.
\bibitem{DR} S. Dittmaier, M. Roth, hep-ph/0206070.
\bibitem{KZ} K.~Ko\l odziej, M.~Zra\l ek, Phys. Rev. D43 (1991) 3619.
\bibitem{JK} F. Jegerlehner, K. Ko\l odziej, Eur. Phys. J. C12 (2000) 77.
\bibitem{MC} F.A. Berends. P.H. Daverveldt, R. Kleiss, Nucl. Phys. B253
             (1985) 441;\\
             R. Kleiss, CERN Yellow Report 89-08 (1989) Vol. 3, p. 9, 
             eds. G. Altarelli, R. Kleiss and C. Verzegnassi;\\
             J. Hilgart, R. Kleiss, F. Le Diberder, Comput. Phys. Commun. 75
             (1993 ) 191.
             
\end{thebibliography}
\end{document}